\documentclass{aa}
\usepackage{graphicx}
\usepackage{natbib}
\usepackage{txfonts}
\bibpunct{(}{)}{;}{a}{}{,} % to follow the A&A style

\def\ls{LS~5039}

\def\j0632{HESS~J0632$+$057}

\begin{document}

\title{Modeling the three-dimensional pair cascade in binaries}
\subtitle{Application to LS~5039}
\titlerunning{Modeling the three-dimensional pair cascade in binaries. Application to LS 5039}

\author{%author1\inst{1,2}
B. Cerutti \inst{1}
\and J. Malzac \inst{2}
\and G. Dubus \inst{1}
\and G. Henri \inst{1}
}
\authorrunning{B. Cerutti et al.}
\institute{
Laboratoire d'Astrophysique de Grenoble, UMR 5571 CNRS, Universit\'e
Joseph Fourier, BP 53, 38041 Grenoble, France 
\and
Centre d'Etude Spatiale des Rayonnements, OMP, UPS, CNRS, 9 Avenue du Colonel Roche, BP 44346, 31028 Toulouse C\'edex 4, France}

\date{Draft \today}
\abstract
%Context
{LS~5039 is a Galactic binary system emitting high and very-high
  energy gamma rays. The gamma-ray flux is modulated on the orbital
  period and the TeV lightcurve shaped by photon-photon
  annihilation. The observed very-high energy modulation can be
  reproduced with a simple leptonic model but fails to explain the
  flux detected by HESS at superior conjunction, where gamma rays
  are fully absorbed.}
%Aims
{The contribution from an electron-positron pair cascade could be
  strong and prevail over the primary flux at superior conjunction. The created pairs can be isotropized by the magnetic field, resulting in a three-dimensional cascade. The aim of this article is to investigate the gamma-ray radiation from this pair cascade in
  LS~5039. This additional component could account for HESS
  observations at superior conjunction in the system.}
%Methods
{A semi-analytical and a Monte Carlo method for computing three-dimensional cascade radiation are presented and applied in the
  context of binaries. The cascade is decomposed into discrete
  generations of particles where electron-positron pairs are assumed
  to be confined at their site of creation. Both methods give similar
  results. The Monte Carlo approach remains best suited to calculation of a multi-generation cascade.}
%Results
{Three-dimensional cascade radiation contributes significantly at every orbital phase in the TeV lightcurve, and dominates close to superior conjunction. The amplitude of the gamma-ray modulation is correctly reproduced for an inclination of the orbit of $\approx 40\degr$. Primary pairs should be injected close to the compact object location, otherwise the shape of the modulation is not explained. In addition, synchrotron emission from the cascade in X-rays constrains the ambient magnetic field to below 10~G.}
%Conclusion
{The radiation from a three-dimensional pair cascade can account for the TeV flux detected by HESS at superior conjunction in LS~5039, but the very-high energy spectrum at low fluxes remains difficult to explain in this model.}

\maketitle

\section{Introduction}

LS~5039 was first identified as a high-mass X-ray binary by
\citet{1997A&A...323..853M}. This binary system is composed of a
massive O type star and an unknown compact object, possibly a young
rotation-powered pulsar
\citep{2005A&A...430..245M,2006A&A...456..801D}. LS~5039 was detected
as a very high-energy ($>100$~GeV, VHE) gamma-ray source by HESS
\citep{2005Sci...309..746A} modulated on the orbital period
\citep{2006A&A...460..743A}. In a leptonic scenario, the gamma-ray emission is produced by inverse Compton scattering of stellar photons on energetic electron-positron pairs injected and accelerated by a rotation-powered pulsar (pulsar wind nebula scenario) or in a relativistic jet powered by accretion on the compact object (microquasar scenario). Most of the VHE modulation is probably caused by absorption of gamma rays in the intense UV stellar radiation
field set by the massive star
\citep{2005ApJ...634L..81B,2006MNRAS.368..579B,2006A&A...451....9D}.

Pairs produced in the system can upscatter a substantial fraction of the
absorbed energy into a new generation of gamma rays and initiate a
cascade of pairs. The radiation from the full cascade can
significantly increase the transparency of the source, particularly at
orbital phases where the gamma-ray opacity is high
($\tau_{\gamma\gamma}\gg1$). A one-zone leptonic model applied to LS~5039 explains the lightcurve and the spectral features at VHE \citep{2008A&A...477..691D}, and yet, this model cannot account for the flux detected by HESS at superior conjunction where gamma rays should be fully absorbed. Pair cascading was mentioned as a possible solution for this disagreement \citep{2006A&A...460..743A}.

The development of a cascade of pairs depends on the ambient magnetic
field intensity. If the magnetic deviations on pair trajectories can
be neglected, the cascade grows along the line joining the source to
the observer. The cascade is one-dimensional. In this case, the cascade contribution is too strong close to superior conjunction in LS~5039. A one-dimensional cascade can be ruled out by HESS
observations \citep{2009A&A...507.1217C} (see the model in
\citealt{2008APh....30..239S} for an alternative solution). If the magnetic field is strong enough to deviate and confine electrons in the system, pairs radiate in all directions and a three-dimensional cascade is initiated \citep{1997A&A...322..523B}. The development of a three-dimensional cascade in LS~5039 is possible and was investigated by \citet{2006MNRAS.368..579B,2007A&A...464..259B} with a Monte Carlo method and by \citet{2008A&A...482..397B} with a semi-analytical method.

\citet{2008A&A...482..397B} derived the non thermal emission produced by the first generation of pairs in gamma-ray binaries. In their model, the density of secondary pairs is averaged over angles describing the mean behavior of the radiating pairs in the system. Here, we aim to investigate the detailed angular dependence in the gamma-ray emission from pairs in the cascade. In the microquasar scenario, \citet{2007A&A...464..259B} finds consistent flux at superior conjunction in LS~5039 if the emission originates farther along the jet ($>10~R_{\star}$) whose direction is assumed to be perpendicular to the orbital plane, including the synchrotron losses. The role of three-dimensional cascade is revisited here in the pulsar wind nebula scenario \citep{1981MNRAS.194P...1M,2006A&A...456..801D}, where the VHE emitter is close to the compact object location. The aim of this article is to corroborate HESS observations of LS~5039 and to constrain the ambient magnetic field strength in the system, using a semi-analytical and a Monte Carlo computation methods. The Monte Carlo code used in the following was previously applied to the system Cygnus~X-1 for similar reasons \citep{2009MNRAS.tmpL.175Z}.

The paper is divided as follows. Sect.~2 gives the main conditions to
initiate a three-dimensional cascade in \ls. The semi-analytical
approach and the Monte Carlo code for cascading calculations are
presented in Sect.~3 and the main features of a three-dimensional pair
cascade in binaries are discussed in Sect.~4. Sect.~5 is dedicated to
the full calculation of a three-dimensional cascade in \ls. The effect
of the ambient magnetic field intensity is also investigated in this
part. The conclusions of the article are exposed in the last section.

In the following, we use the term ``electrons'' to refer indifferently to electrons and positrons.

\begin{figure}
\centering
\resizebox{\hsize}{!}{\includegraphics{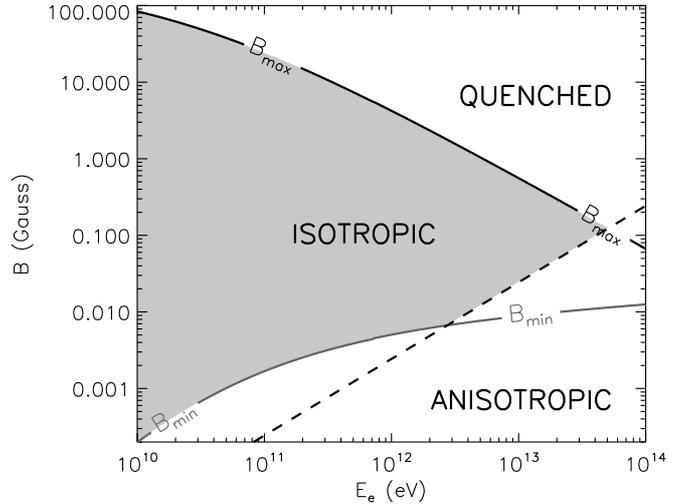}} 
  \caption{This map shows the domain (gray surface,`ISOTROPIC') where
    a three-dimensional isotropic cascade can be initiated as a
    function of the ambient magnetic field $B$ and the energy of the
    electron $E_e$. This calculation is applied to \ls\ at periastron
    (orbital separation $d\approx 0.1$~AU). The upper-limit is bounded
    by the black solid line labeled `$B_{max}$' and the lower-limit
    by the gray solid line `$B_{min}$'. For $B>B_{max}$ (`QUENCHED'),
    synchrotron losses dominate and the cascade is inhibited. For
    $B<B_{min}$ (`ANISOTROPIC') the cascade is not locally isotropized
    and depends on the magnetic field structure. The isotropic domain
    is truncated at VHE as the pairs escape from the system (below the
    dashed line).}
\label{domain}
\end{figure}

\section{The magnetic field for 3D cascade}

The development of the cascade is dictated by the intensity of the ambient magnetic field in the binary environment. The main conditions for the existence of a three-dimensional cascades have been investigated by \citet{1997A&A...322..523B} and are reviewed here and applied to \ls.

The magnetic field $B$ must be high enough to locally isotropize pairs once created. This condition is fulfilled if the Larmor radius of the pair $R_L$ is shorter than the inverse Compton energy losses length given by $\lambda_{cool}= -\beta_e c \gamma_e/\dot{\gamma}_e$, where $\gamma_e=1/(1-\beta_e^2)^{1/2}$ is the Lorentz factor of the electron and $\dot{\gamma}_e\equiv d\gamma_e/dt$ is the Compton energy losses. This provides a lower-limit for the magnetic field. In the Thomson regime, this is given by
%\begin{equation}
%B_{T}\ga 8\times 10^{-5}~\gamma_3 T_{\star,4}^3 R^2_{\star,10} d^{-2}_{0.1}~\rm{G},
%\label{bmin_th}
%\end{equation}
\begin{equation}
B_{T}\ga 2\times 10^{-6}~\gamma^2_3 T_{\star,4}^4 R^2_{\star,10} d^{-2}_{0.1}~\rm{G},
\label{bmin_th}
\end{equation}
writing $\gamma_3=\gamma_e/10^3$, $T_{\star,4}=T_{\star}/40~000$~K and $R_{\star,10}=R_{\star}/{10~R_{\odot}}$ the temperature and radius of the companion star, and $d_{0.1}=d/0.1$~AU the orbital separation. Using the approximate formula for Compton energy losses \citep{1970RvMP...42..237B}, the same condition in the extreme Klein-Nishina regime holds if
\begin{equation}
B_{KN}\ga 1.6\times 10^{-3}~T_{\star,4}^2 R^2_{\star,10} d^{-2}_{0.1}\left[\ln\left(\gamma_6 T_{\star,4}\right)+2.46\right]~\rm{G}.
\label{bmin_kn}
\end{equation}
If the Larmor radius is compared with the Compton mean free path given by $\lambda_{ic}\sim 1/n_{\star}\sigma_{ic}$, where $n_{\star}$ is the stellar photon density and $\sigma_{ic}$ the Compton cross section, the condition on the magnetic field is more restrictive. In the Thomson regime, the electron loses only a small fraction of its total energy per interaction, hence $\lambda_{cool}>\lambda_{ic}$. In the Klein-Nishina regime, most of the electron energy is lost in a single scattering and $\lambda_{cool}\approx\lambda_{ic}$. Because the cascade occurs mostly in the Klein-Nishina regime in gamma-ray binaries, both conditions lead approximatively to the same lower limit for the ambient magnetic field.

In addition to this condition, pairs are assumed to be isotropized at their creation site for simplicity. Pairs will be randomized if the ambient magnetic field is disorganized. Isotropization of pairs in the cascade will also occur due to pitch angle scattering if the magnetic turbulence timescale is smaller than the energy loss timescale ({\em e.g.} if it is on the order of the Larmor timescale). For lower magnetic field intensity (`anisotropic' domain in Fig.~\ref{domain}), the cascade remains three-dimensional but then pairs cannot be considered as locally isotropized. In this case, the trajectories of the particles should be properly computed as in {\em e.g.} \citealt{2005MNRAS.356..711S}. For $B\la 10^{-8}~$G, the cascade is one-dimensional \citep{2009A&A...507.1217C}.

If the magnetic field is too strong, pairs locally isotropize but cool down via synchrotron radiation rather than by inverse Compton scattering. Most of the energy is then emitted in X-rays and soft gamma rays, {\em i.e.} below the threshold energy for pair production. The cascade is quenched as soon as the first generation of pairs is produced. This condition gives an upper-limit for the magnetic field. Synchrotron losses are smaller than inverse Compton losses $\dot{E}_{syn}<\dot{E}_{ic}$ for
\begin{equation}
B_{T}\la 163~T_{\star,4}^{2} R_{\star,10}^{} d_{0.1}^{-1}~\rm{G},
\label{bmax_th}
\end{equation}
in the Thomson regime and for
\begin{equation}
B_{KN}\la 4.7~\gamma_6^{-1}T_{\star,4}^{}R_{\star,10}^{}d^{-1}_{0.1}\left[\ln\left(\gamma_6 T_{\star,4}\right)+2.46\right]^{1/2}~\rm{G}
\label{bmax_kn}
\end{equation}
in the deep Klein-Nishina regime \citep{1970RvMP...42..237B}. It can be noticed that the most relevant upper-limit for the magnetic field strength is given by the Thomson formula in Eq.~(\ref{bmax_th}), since high-energy particles ($E_e\ga 1$~GeV) with $B_{KN}<B<B_T$ can cool down and get into the cascade domain.

Figure~\ref{domain} shows the complete domain where a three-dimensional `isotropic' cascade can be initiated in \ls, combining the lower and upper-limit for $B$. This domain encompasses plausible values for the ambient magnetic field in the system. It is worthwhile to note that for very high-energy electrons $E_e\ga 45~d_{0.1}B_{0.1}$~TeV, where $B_{0.1}=B/0.1~$G, the Larmor radius becomes greater than the binary separation in \ls~(Fig.~\ref{domain}). In this case, the local magnetic confinement approximation of particles is not appropriate anymore. This is unlikely to happen in \ls\ if the VHE emission has a leptonic origin since HESS observations shows an energy cut-off for photons at $\approx 10$~TeV.

\begin{figure}
\centering
\resizebox{\hsize}{!}{\includegraphics{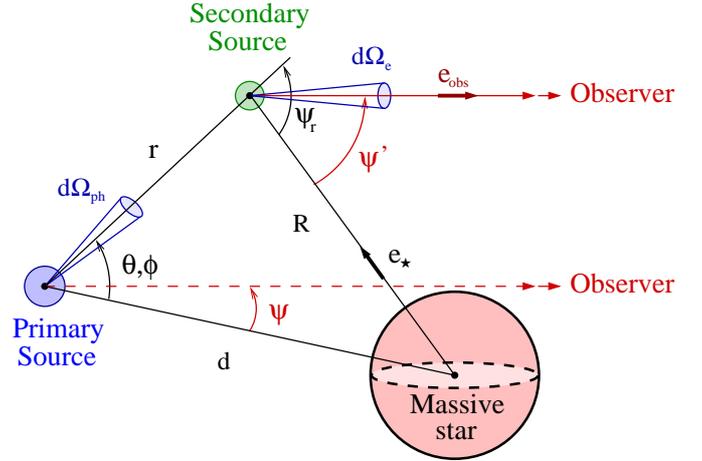}} 
  \caption{In this figure is depicted the geometric quantities useful for three-dimensional pair cascading calculation in $\gamma$-ray binaries. The primary source is point-like and coincides with the compact object location. The system is viewed at an angle $\psi$ by a distant observer. The absorption of primary gamma rays at the distance $r$ in the $(\theta,\phi)$ direction creates a secondary source of radiation, viewed at an angle $\psi'$ by the observer.}
\label{geo}
\end{figure}

\section{Computing methods}

Contrary to the one-dimensional case, three-dimensional pair cascading cannot be explicitly computed. Nevertheless, it is possible to decompose the cascade into successive generations of particles. Two different approaches are presented below, one based on semi-analytical calculations and the other on a Monte Carlo code. In both models, the primary source of gamma rays is point-like and coincident with the compact object position as it is depicted in Fig.~\ref{geo}. The origin and the angular dependence of the primary gamma-ray flux are not specified at this stage. These methods are general and could be applied to any other astrophysical context involving 3D pair cascading.

\begin{figure*}
\resizebox{17cm}{!}
{\includegraphics*{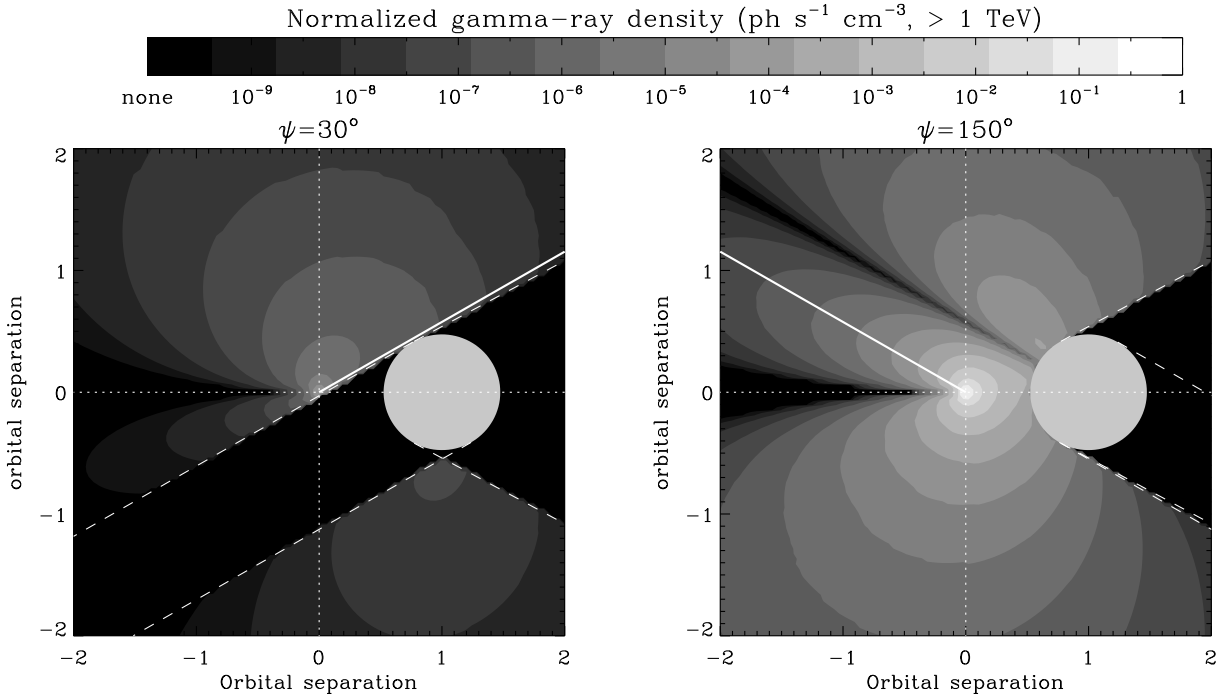}}
  \caption{Spatial distribution of the escaping ({\em i.e.} including the effect of gamma-ray absorption) VHE photon density (ph $\rm{s^{-1}}~\rm{cm^{-3}}$) emitted by the first generation of electrons (isotropized) in the cascade as observed by a distant observer in LS~5039 at superior ({\em left}) and inferior ({\em right}) conjunction. These maps show the gamma-ray density in logarithmic scale (common for both maps), where bright and dark regions correspond respectively to high and low density. Each map is a slice of the 3D gamma-ray emission distribution in the plane that contains the observer (whose direction is indicated by the white solid line) and both stars, computed with the semi-analytical method. The primary source of gamma rays is isotropic and lies at the compact object location (origin). White dashed lines delimit the eclipsed regions (for the primary source and the observer) by the massive star (bright uniform disk). The massive star is assumed point like and mono-energetic in the calculations of radiative processes. Distances are normalized to the orbital separation.}
\label{map}
\end{figure*}

\subsection{Semi-analytical}

A beam of primary gamma rays propagating in the direction defined by the spherical angles $\theta$ and $\phi$ (see Fig.~\ref{geo}), produces at a distance $r$ to the primary source the first generation of pairs.  In the point-like and mono-energetic star approximation, the density of electrons and positrons injected per unit of time, energy and volume ($\rm{s}^{-1}\rm{erg}^{-1}\rm{cm}^{-3}$) is
\begin{equation}
\frac{dN_e^{(1)}}{dt dE_e d\mathcal{V}}=2\int_{\epsilon_1}\frac{1}{r^2}\frac{dN_{ph}^{(0)}}{dtd\epsilon_1 d\Omega_{ph}} g_{\gamma\gamma}~e^{-\tau_{\gamma\gamma}(r)}d\epsilon_1,
\label{pair_inj}
\end{equation}
where $dN_{ph}^{(0)}/dtd\epsilon_1 d\Omega_{ph}$ is the density of primary gamma rays of energy $\epsilon_1$, $g_{\gamma\gamma}$ the anisotropic pair production kernel \citep{1971MNRAS.152...21B,1997A&A...325..866B,2009A&A...507.1217C} and $\tau_{\gamma\gamma}(r)$ the $\gamma\gamma$-opacity integrated from the source to the position $r$. This new density of pairs is spatially extended and anisotropic but is symmetric with respect to the line joining the star to the primary source. For a fixed stellar radiation field and a given steady source of primary gamma rays, pair production provides a continuous source of fresh electrons injected in the binary system environment.

Pairs are supposed to be immediately confined and isotropized by the local magnetic field at their creation site. The binary vicinity is surrounded by a plasma of isotropic pairs cooling {\em via} synchrotron radiation and inverse Compton scattering. For simplicity, electrons are assumed to have enough time to radiate before escaping their site of injection and the advection of particles by the massive star wind is ignored although this can have some impact \citep{2008A&A...482..397B}. For a 1~TeV electron, the radiative cooling timescales in LS~5039 are $t_{ic}\approx 20~$s (inverse Compton, at the compact object location) and $t_{syn}\approx 400$~s (synchrotron, for $B=1~$G). The maximum escaping timescale is given by the advection time of pairs by the stellar wind. Taking a wind terminal velocity $v_\infty\approx 2400$~km~$\rm{s}^{-1}$ for the massive star in LS~5039 \citep{2004ApJ...600..927M}, $t_{esc}=d/v_\infty\approx 6\times10^3$~s~$\gg t_{ic}$ and $t_{syn}$. In the case where pairs would escape the system at the speed of light, electrons have just enough time to radiate by inverse Compton scattering ($t_{esc}=d/c\approx 50~$s $\la t_{ic}$). This extreme situation is unlikely since pairs are confined by the ambient magnetic field but provides a lower limit for the escaping timescale in the system. Assuming that $t_{esc}\gg t_{ic}$ and $t_{syn}$ is a rather good approximation in LS~5039 for the high-energy particles.

The steady-state particle distribution in $\rm{erg}^{-1}\rm{cm}^{-3}\rm{sr}^{-1}$ is \citep{1964ocr..book.....G}
\begin{equation}
\frac{dN_e^{(1)}}{dE_e d\mathcal{V}d\Omega_e}=\frac{1}{\left|\dot{E}_e\right|}\int_{E_e}^{+\infty}\frac{1}{4\pi}\frac{dN_{e}^{(1)}}{dtdE'_e d\mathcal{V}} dE'_e,
\label{cool}
\end{equation}
with $\dot{E_e}=\dot{E}_{ic}+\dot{E}_{syn}$ the inverse Compton and synchrotron losses and $\mathcal{V}$ the volume encircling the binary. Note that the annihilation of pairs is not considered in this calculation since this effect would be important only for pairs that are almost thermalized. Triplet pair production $\gamma+e^{\pm}\rightarrow e^{\pm}+e^{+}+e^{-}$ (see {\em e.g.} \citealt{1991MNRAS.253..235M}) is ignored too (see the discussion in \citealt{2009A&A...507.1217C}, Sect.~2.1).

The total inverse Compton radiation produced by the first generation of pairs observed by a distant observer is given by
\begin{equation}
\frac{dN_{ic}^{(1)}}{dt d\epsilon_1 d\Omega_e}=\iint \frac{dN_e^{(1)}}{dE_e d\mathcal{V} d\Omega_e} n_{\star} \frac{dN_{ic}}{dtd\epsilon_1} e^{-\tau_{\gamma\gamma}} dE_e d\mathcal{V},
\label{compton}
\end{equation}
where $n_{\star}$ is the stellar photon density in $\rm{cm}^{-3}$, $dN_{ic}/dtd\epsilon_1$ the anisotropic inverse Compton kernel \citep{2008A&A...477..691D} and $\tau_{\gamma\gamma}$ the absorption from the secondary source up to the observer. Depending on the relative position of the secondary source, the massive star and the observer, inverse Compton emission is anisotropic though pairs are isotropic. The secondary source is seen at an angle $\psi'$ with $\cos\psi'=-\vec{e}_{\star}\cdot\vec{e}_{obs}$ (Fig.~\ref{geo}) so that
\begin{equation}
\cos\psi'=\cos\psi\cos\left(\psi_r-\theta\right)-\sin\psi\sin\left(\psi_r-\theta\right)\cos\phi.
\label{cos_psi}
\end{equation}
In the point-like star approximation, this viewing angle $\psi'$ is
related to the interaction angle $\theta_0$ between photons and
electrons such as $\cos\psi'=-\cos\theta_0$. Similarly to inverse Compton scattering, the total synchrotron radiation produced by the first generation of pairs is
\begin{equation}
\frac{dN_{syn}^{(1)}}{dt d\epsilon_1 d\Omega_e}=\iint \frac{dN_e^{(1)}}{dE_e d\mathcal{V} d\Omega_e}\frac{dN_{syn}}{dtd\epsilon_1} dE_e d\mathcal{V},
\label{synchrotron}
\end{equation}
with $dN_{syn}/dtd\epsilon_1$ the synchrotron kernel averaged over an
isotropic distribution of pitch angles to the magnetic field (see {\em
e.g.} \citealt{1970RvMP...42..237B}).

This semi-analytical method can be extended to an arbitrary
number of generations. By replacing the primary density of gamma rays
in Eq.~(\ref{pair_inj}) by the new density of created photons
Eqs.~(\ref{compton})-(\ref{synchrotron}), the second generation of pairs and gamma-rays in the cascade can be computed, and so on for the next generations.

\begin{figure}
\centering
\resizebox{\hsize}{!}{\includegraphics{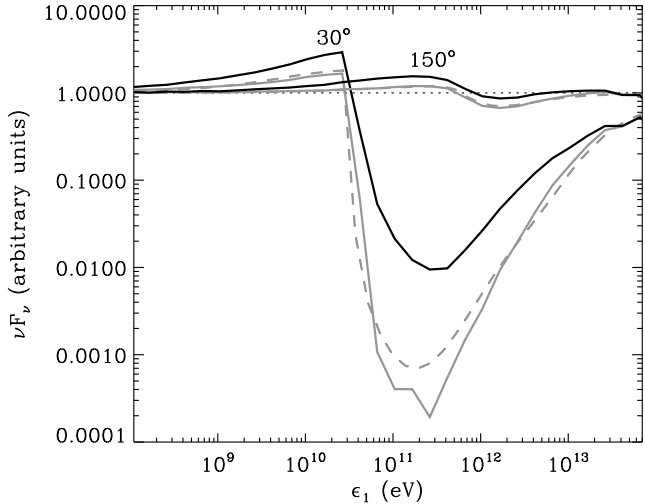}} 
  \caption{The full cascade radiation (all generations) computed with the Monte Carlo code (black solid lines) and the primary injected gamma-ray source (isotropic, dotted line) are shown for $\psi=30\degr$ and $150\degr$. The Monte Carlo output (solid gray lines) is compared with the semi-analytical calculations (dashed gray lines) in the one-generation cascade approximation. There is no magnetic field but pairs are still assumed to be confined and isotropized. The massive star is point like and mono-energetic.}
\label{compare}
\end{figure}

\subsection{Monte Carlo}

We also used a Monte-Carlo code to simulate the development of the full electromagnetic pair cascade in the radiation field of the star. In this calculation the path and successive interactions of photons and leptons are tracked until they escape the system (in practice until they reach a distance about 10 times the binary separation). This code was previously used by \citet{2009MNRAS.tmpL.175Z} to model the TeV emission of Cygnus X-1. It is similar in scope and capabilities to the code of \citet{1997A&A...322..523B}. The present code was developped completely independently, and most of the random number generation techniques used for computing photon path and simulating the interactions are very different from those used by Bednarek. Perhaps the most important difference is that the Compton interactions are simulated without any approximation, even in the deep Klein-Nishina regime. Also, in order to reduce the computing time required to achieve high accuracy at high energies, we use a weighting technique which avoids  following every particle of the cascade down to low energies. The results of both codes were compared and found compatible \citep{2009MNRAS.tmpL.175Z}.

\section{Three-dimensional pair cascade radiation}

For illustrative purpose only, the primary source of gamma rays is assumed isotropic in this section. This assumption allows a better appreciation of the intrinsic anisotropic effects of the pair cascade emission in binaries. Primary gamma rays are injected with a $-2$ (photon index) power-law spectrum at the location of the compact object. For simplicity, the massive star is assumed here point-like and mono-energetic. More realistic assumptions (injection of isotropic electrons, black body and finite size companion star) are considered for the calculation of the 3D cascade emission in LS~5039 in the next Section (Sect.~\ref{sect_ls}).

\subsection{Spatial distribution of gamma rays in the cascade}{\label{spatial}}

Figure~\ref{map} shows the spatial distribution of the first generation of escaping TeV gamma rays seen by a distant observer ({\em i.e.} including the effect of gamma-ray absorption) produced by the cascade in LS~5039 at both conjunctions (for an inclination of the orbit $i=60\degr$). These maps are computed with the semi-analytical approach. The massive star is assumed point-like for the computation of radiative processes but eclipses are considered. No pairs can be created behind the star with respect to the primary source of gamma rays. Also, gamma rays produced behind the star with respect to the observer are excluded from the overall cascade radiation (see black regions in Fig.~\ref{map}). Synchrotron radiation is neglected in this part: pairs radiate only {\em via} inverse Compton scattering.

The spatial distribution of gamma rays is extended and is not rotationally symmetric about the line joining the two stars (contrary to pairs) since the observed inverse Compton emission depends on the peculiar orientation of the observer with respect to the binary system. No gamma rays are emitted along the line joining the star to the observer direction (see Fig.~\ref{map}, {\em right panel}) because pairs undergo rear-end collisions with the stellar photons ($\vec{e}_{\star}\cdot\vec{e}_{obs}=1$). This effect is smoothed if the finite size of the massive star is considered. The escaping gamma-ray density at inferior conjunction is more important than at superior conjunction as TeV photons suffer less from absorption.

\begin{figure*}
\resizebox{17cm}{!}
{\includegraphics{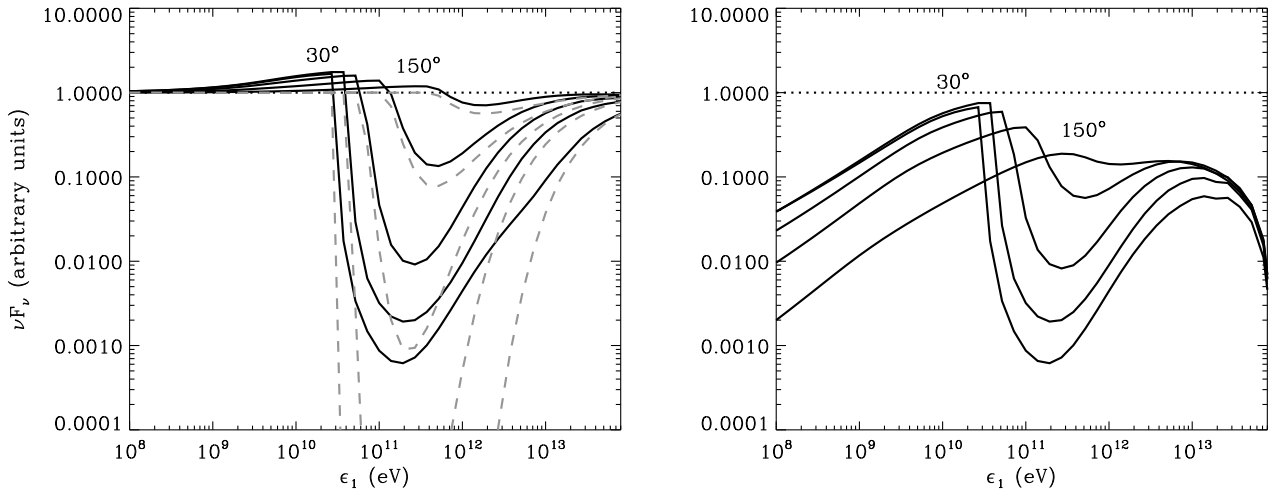}} 
  \caption{Cascade radiation emitted by the first generation computed
    with the semi-analytical method in \ls\ at periastron for
    $\psi=30\degr,~60\degr,~90\degr,~120\degr$ and $150\degr$. {\em Left:} The escaping gamma-ray spectrum (solid line) is compared to the pure-absorbed (dashed gray line) and injected (isotropic, dotted line) spectra. The radiation from the cascade only is shown on the {\em right panel}. Synchrotron radiation is ignored and the massive star is point like and mono-energetic. }
\label{anis}
\end{figure*}

\subsection{One and multi-generation cascade}

The semi-analytical method is ideal to study the first generation of particles in the cascade as it provides quick and accurate solutions. In principle, this method can be extended to an arbitrary number of generation but the computing time increases tremendously. The Monte Carlo approach is well suited to treat complex three dimensional radiative transfer problems. With this method, the full cascade radiation (including all generations) can be computed with a reasonable amount of time but a large number of events is required to have enough statistics for accurate predictions.

Figure~\ref{compare} gives the escaping gamma-ray spectra at both conjunctions in LS~5039. The Monte Carlo output is compared with the semi-analytical results in the same configuration as in Fig.~\ref{map} for $\psi=30\degr$ and $150\degr$. Both approaches give similar results for the first generation of gamma rays. There are slight differences mainly due to statistical and binning effect in the Monte Carlo result, particularly at $\psi=30\degr$ where the absorption is high. The contribution from additional generations of pairs to the cascade radiation is of major importance as it dominates the overall escaping gamma-ray flux where the primary photons are fully absorbed. The Monte Carlo approach is needed to compute the cascade radiation where absorption is strong {\em i.e.} at superior conjunction. In practice, the one-generation approximation catches the main features of the full three-dimensional pair cascade calculation elsewhere along the orbit.

\subsection{Comparison with one-dimensional cascade}{\label{1d3d}}

Three-dimensional cascade radiation presents identical spectral features to the one-dimensional limit \citep{2009A&A...507.1217C} (Fig.~\ref{anis}). Below the threshold energy for pair production, {\em i.e.} $\epsilon_1<m^2_e c^4/2\epsilon_0\left(1-\cos\theta_0\right)$ with $\epsilon_0$ the stellar photon energy, pairs cool down {\em via} inverse Compton scattering in the Thomson regime and accumulate at lower energy in a $\sim-1.5$ photon index power-law tail. Above, emission and absorption compete, giving rise to a dip in the spectrum. At higher energies ($\epsilon_1\ga 10~$TeV), the gamma-ray production in the cascade declines due to Klein-Nishina effect in inverse Compton scattering and pair production becomes less efficient.

\begin{figure}
\centering
\resizebox{\hsize}{!}{\includegraphics{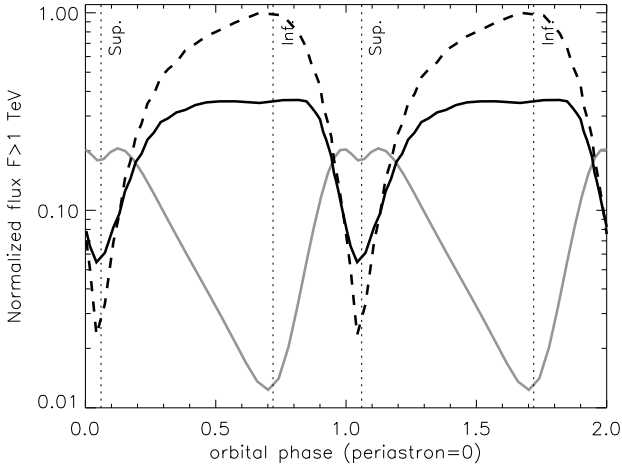}}
  \caption{Modulation of the TeV flux produced by a three-dimensional (Monte Carlo calculation, black solid line) and one-dimensional (semi-analytical calculation see \citealt{2009A&A...507.1217C}, gray solid line) cascade in LS~5039 as a function of the orbital phase (two full orbits). Synchrotron radiation is ignored for the computation of 3D cascade radiation. The primary absorbed flux (identical injection as in Fig.~\ref{anis}, {\em i.e.} isotropic) is shown (dashed line) for comparison. Conjunctions are indicated by vertical dotted lines. Orbital parameters are taken from \citet{2005MNRAS.364..899C} for an inclination $i=60\degr$. The companion star is point like and mono-energetic.}
\label{mod_cas}
\end{figure}

Three-dimensional cascade radiation has a strong angular dependence (Fig.~\ref{anis}) that differs significantly from the one-dimensional case. Figure~\ref{mod_cas} presents the modulation of the TeV radiation from a 1D and 3D cascade along the orbit in LS~5039 (the one-dimensional cascade radiation is calculated with the method described in \citealt{2009A&A...507.1217C}). \citet{2006MNRAS.368..579B} found a similar modulation for the 3D cascade radiation. Both contributions are anti-correlated. Contrary to the one-dimensional cascade, the three-dimensional cascade radiation preserves the modulation of the primary absorbed source of gamma rays since pairs do not propagate. Peaks and dips remain at conjunctions. In both cases, the cascade radiation flux prevails at superior conjunction where the primary flux is highly absorbed. Note that a small dip in the 1D cascade radiation appears at superior conjunction because absorption slightly dominates over emission. The 3D cascade contributes less (by a factor $\approx 3$) than the 1D cascade to the total TeV flux at this orbital phase.

\subsection{The effect of the ambient magnetic field}{\label{sect_mag}}

\begin{figure}
\centering
\resizebox{\hsize}{!}{\includegraphics{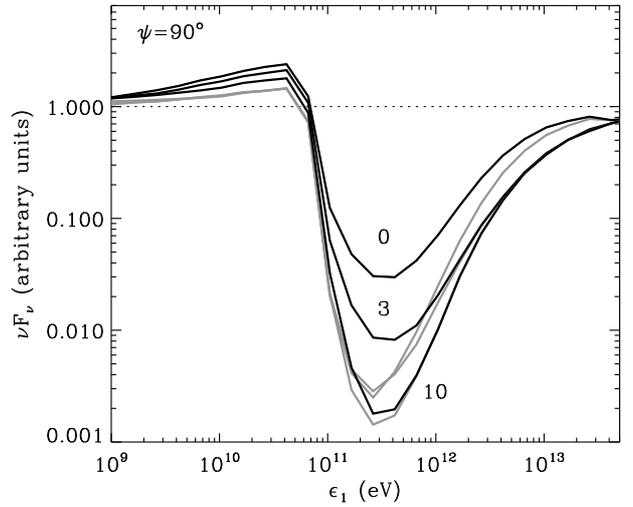}} 
  \caption{Effect of the ambient magnetic field on the cascade radiation. The cascade is computed with the same parameters (Monte Carlo approach) as used in Fig.~\ref{compare} for $\psi=90\degr$ with an uniform magnetic field $B=0$ (top), 3 and $10~$G (bottom). The full escaping gamma-ray spectra (all generations, black lines) is compared with the one-generation approximation (gray lines) and the injected isotropic spectra (dotted line). The companion star is point like and mono-energetic.}
\label{mag}
\end{figure}

Synchrotron radiation has a significant impact on the cascade spectrum. Figure~\ref{mag} shows the effects of an uniform ambient magnetic field on the cascade radiation for $B=0$, 3 and 10~G. The VHE emission is quenched as synchrotron radiation becomes the dominant cooling channel for electrons produced in the cascade ($t_{ic}>t_{syn}$). The large contribution of the cascade in the TeV band is preserved if the magnetic field does not exceed a few Gauss (see Fig.~\ref{domain}). Synchrotron radiation contributes to the total flux in the X-ray to soft gamma-ray energy band. These photons do not participate to the cascade as their energy does not exceed 100~MeV, which is insufficient for pair production with the stellar photons.

Figure~\ref{mag} compares also the contribution from the first generation of gamma rays with the full cascade radiation. For low magnetic field ($B\la5~$G), all generations should be considered in the calculation. For higher magnetic field ($B>5~$G), the first generation of gamma rays dominates the total cascade radiation. Only a few pairs can radiate beyond the threshold energy for pair production and the cascade is quenched.

A non-uniform magnetic field was also investigated for a toroidal or dipolar magnetic structure generated by the massive star ({\em i.e.} with a $R^{-1}$ or $R^{-3}$ dependence). These configurations do not give different results compared with the uniform case. Most of the cascade radiation is produced close to the primary source (see Sect.~\ref{spatial}) and depends mostly on the magnetic field strength at this location.

\section{Three-dimensional cascades in LS~5039}{\label{sect_ls}}

%\begin{figure}
%\centering
%\resizebox{\hsize}{!}{\includegraphics*{casp.eps}} 
%  \caption{Expected SUPC ($0.45<\phi<0.9$, dashed line) and INFC ($\phi<0.45$ or $\phi>0.9$, solid line) averaged spectra in \ls, where the contribution of the first generation of gamma rays in the cascade is taken into account. Spectra are compared with HESS (red bowties) and EGRET (gray bowtie) observations.}
%\label{casp}
%\end{figure}

The full cascade radiation calculation is applied to LS~5039 and discussed below. The black body spectrum and the spatial extension of the massive star are taken into account in this part. The primary source of gamma rays is computed here following the model described in \citet{2008A&A...477..691D} where the pulsar is assumed to inject energetic electron-positron pairs with an isotropic power-law energy distribution at the shock front, expected to lie at the vicinity of the compact star. Taking $v_{\infty}=2400~\rm{km~s}^{-1}$, $\dot{M}=10^{-7}~M_{\odot}~\rm{yr}^{-1}$ for the massive star wind \citep{2004ApJ...600..927M}, and a pulsar spin-down luminosity $L_p=10^{36}~\rm{erg~s}^{-1}$, both wind momenta balance at a distance $r_{shock}\sim 0.1 d$ from the pulsar. Pairs generated by the pulsar emit {\em via} inverse Compton scattering on stellar photons the primary gamma-ray photons. Contrary to the previous section, the primary gamma-ray source is highly anisotropic. The orbital parameters of the system are taken from \citet{2005MNRAS.364..899C}. New optical observations of LS~5039 have been carried out recently by \citet{2009ApJ...698..514A} where slight corrections to the orbital parameters have been reported, but these do not change the results below.

\subsection{TeV orbital modulation}

\begin{figure}
\centering
%\resizebox{\hsize}{!}{
\includegraphics*[width=8.5cm]{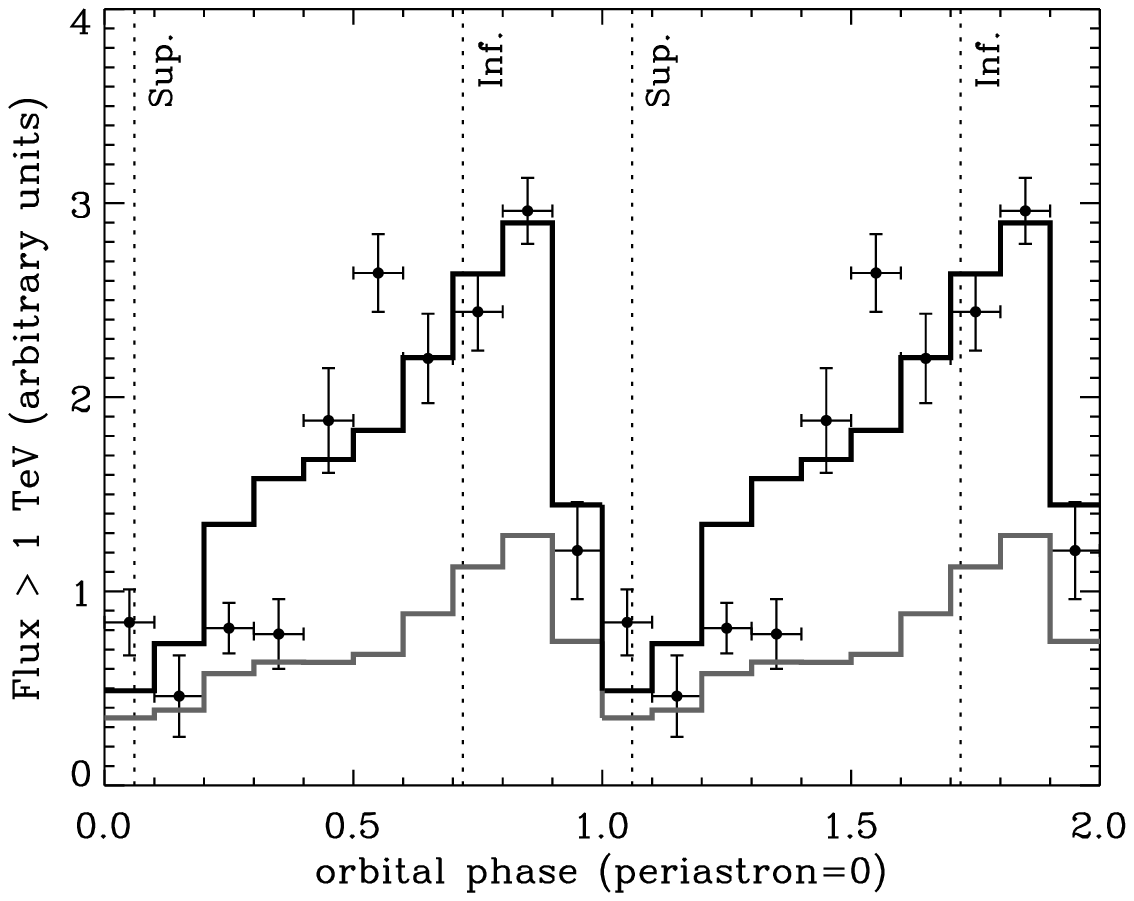}
\includegraphics*[width=8.5cm]{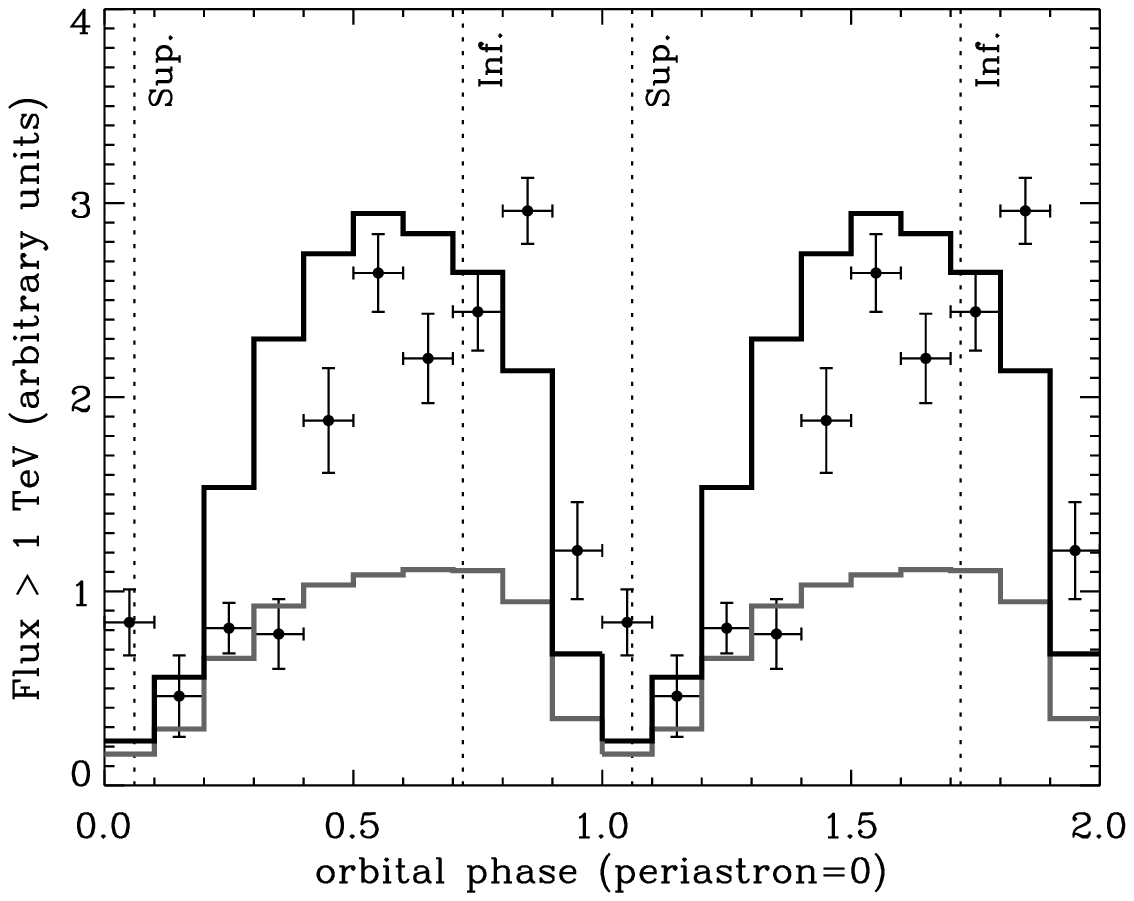}%}
  \caption{Theoretical integrated flux above 1~TeV (black solid line) in LS~5039 as a function of the orbital phase (two orbits) with an inclination of the orbit $i=40\degr$ in both panels. The cascade radiation contribution (gray solid line) is computed with the Monte Carlo approach for a constant injection of energy in cooled particles ({\em top}) and for a constant injection of pairs ({\em bottom}) along the orbit. The black-body spectrum and the finite size of the companion star are taken into account. The ambient magnetic field is small ($B<1~$G). Theoretical lightcurves are binned in phase interval of width $\Delta\phi=0.1$ in order to compare with HESS observations (data points) taken from \citet{2006A&A...460..743A}. Both conjunctions (`Sup.' and `Inf.') are indicated with dotted lines.}
\label{light}
\end{figure}

The shape of the TeV light curve can be explained with a one-zone leptonic model \citep{2008A&A...477..691D} that combines emission and absorption. However, it overestimates the amplitude of the modulation (by a factor $\gtrsim 50$ for $i=60^{\circ}$). The TeV flux observed by HESS varies by about a factor 6 with a minimum at the orbital phases $\phi=0.1$-0.2 and a maximum at $\phi=0.8$-0.9 \citep{2006A&A...460..743A}. The radiation from a three-dimensional cascade of pairs decreases the amplitude of the TeV modulation yet conserves the light curve pattern (see Sect.~\ref{1d3d}). The flux remains minimum at superior conjunction ($\phi\approx 0.06$) and maximum just after inferior conjunction ($\phi\approx 0.85$).

The amplitude of the modulation in LS~5039 can be reproduced for an inclination of the orbit $i=40\degr$ (Fig.~\ref{light}, {\em top panel}), assuming a constant energy density of cooled particles along the orbit as in \citet{2008A&A...477..691D}. This assumption imples that the injection of fresh particles depends (roughly) as $d^{-2}$. The ambient magnetic field is $\la 1~$G (if uniform) otherwise emission up to 10~TeV cannot be sustained. For higher inclination ($i\ga 50\degr$), the flux at superior conjunction is too small to explain observations. For lower inclination ($i\la 30\degr$), the amplitude of the light curve becomes too small. If the injection rate of the uncooled primary pairs is instead kept constant along the orbit (Fig.~\ref{light}, {\em bottom panel}), a lower inclination ($i\la30\degr$) is required to reproduce an amplitude consistent with observations. Then, the light curve presents a broad peak centered at $\phi\approx 0.5$. The profile of the modulation is not explained to satisfaction in this case.

\begin{figure}
\centering
\resizebox{\hsize}{!}{\includegraphics*{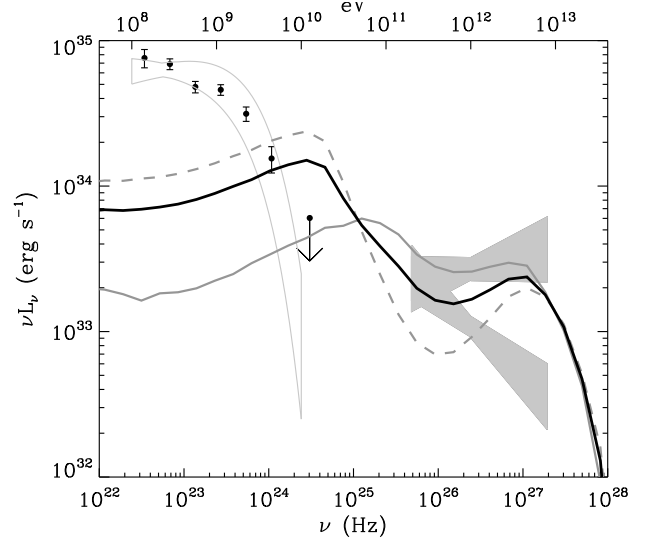}}
  \caption{Theoretical gamma-ray spectum in LS~5039 for `SUPC' ({\em i.e.} averaged over $0.45<\phi<0.9$, gray dashed line) and `INFC' ($\phi<0.45$ or $\phi>0.9$, gray solid line) states as defined in \citet{2006A&A...460..743A} and orbit averaged spectrum (black solid line). Comparison with {\em Fermi} (black data points, \citealt{2009ApJ...706L..56A}) and HESS (red bowties, \citealt{2006A&A...460..743A}) observations.}
\label{spectre_orb}
\end{figure}

The cascade radiation contributes significantly at every orbital phase and dominates the overall gamma-ray flux close to superior conjunction ($0<\phi<0.15$), where the primary flux is highly absorbed. The residual flux observed at superior conjunction is explained by the cascade. The averaged spectra at high and very-high energy are not significantly changed compared with the case without cascade (Fig.~\ref{spectre_orb}, see also Fig.~6 in \citealt{2008A&A...477..691D}). It should be noted that the ratio between the GeV and the TeV flux decreases if a three-dimensional pair cascading is considered. The cascade contributes more at TeV than at GeV energies with respect to the primary source. If spectra are fitted with HESS observations, then the flux expected at GeV energies is too low to explain observations. In addition, this model cannot account for the energy cutoff observed by {\em Fermi} at a few GeV \citep{2009ApJ...706L..56A}. Electrons radiating at GeV and TeV energies may have two different origins. An extra component, possibly from the pulsar itself (magnetospheric or free pulsar wind emission, see \citealt{2009arXiv0912.3722C}) might dominate at GeV energies.

\begin{figure}
\centering
%\resizebox{\hsize}{!}{
\includegraphics*[width=8.5cm]{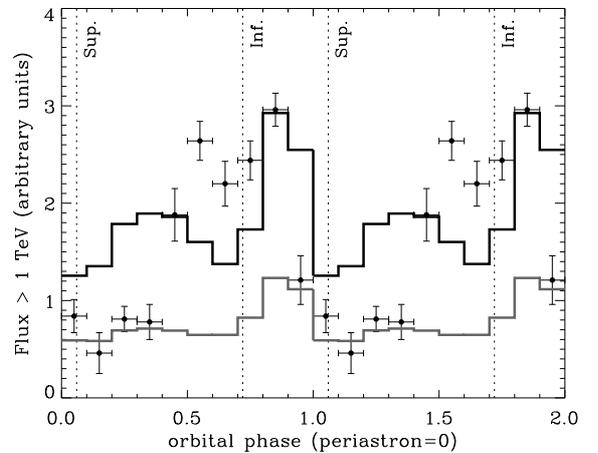}
  \caption{Same as in Fig.~\ref{light} ({\em top panel}) for $i=60^{\circ}$ with a primary source of gamma rays above the compact object and perpendicular to the orbital plane for an altitude $z=2~R_{\star}$.}
\label{light_z}
\end{figure}

\subsection{Constraint on the location of the VHE emitter}

The primary gamma-ray emitter position might not coincide with the compact object location. One possibility is to imagine that particles radiate VHE farther in the orbital plane, for instance backward in a shocked pulsar wind collimated by the massive star wind. In this case, the primary source is less absorbed along the orbit and more power into particles is required to compensate for the decrease of the soft photon density from the companion star. A consistent amplitude could be obtained if the primary gamma rays originate from large distances ($\ga 10~d$), but then the TeV light curve shape is incorrectly reproduced as the tendency for the main peak is to shift towards superior conjunction. Another possibility is to assume that the VHE emitter stands above the orbital plane ({\em e.g.} in a jet). This situation does not differ significantly from the previous alternative. For altitudes $z>2~{R_{\star}}\approx d$, the $\gamma\gamma$-opacity decreases significantly and the escaping VHE gamma-ray flux increases at superior conjunction but the TeV modulation is not reproduced as well (Fig.~\ref{light_z}). Regarding observations, it appears difficult with this model to push the gamma-ray emitter at the outer edge of the system. The primary source should still lie in the vicinity of the compact object ({\em i.e.} at distances smaller than the orbital separation).

\subsection{Constraint on the ambient magnetic field}

\begin{figure}
\centering
\resizebox{\hsize}{!}{\includegraphics{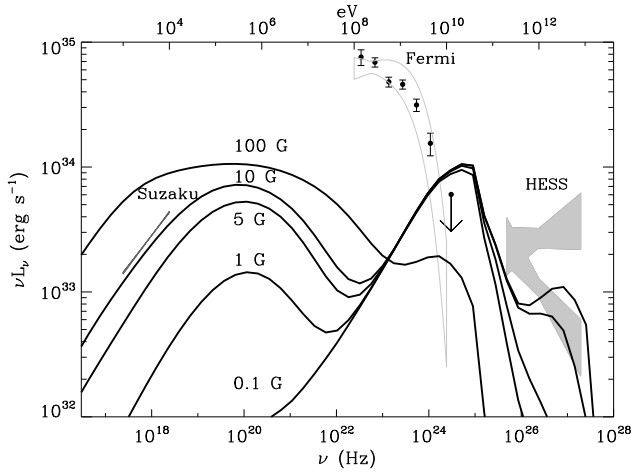}}
  \caption{Orbit averaged spectrum of the first generation of gamma rays in LS~5039 with a uniform magnetic field $B=0.1~$, 1, 5, 10 and 100~G. Comparison with observations from X-rays to TeV energies: {\em Suzaku} \citep{2009ApJ...697..592T}, {\em Fermi} \citep{2009ApJ...706L..56A} and HESS \citep{2006A&A...460..743A} bowties.}
\label{mag_orb}
\end{figure}

The synchrotron radiation produced by secondary pairs can be a dominant contributor to the overall X-ray luminosity as discussed by \citet{2008A&A...482..397B,2008A&A...489L..21B}. Figure~\ref{mag_orb} presents the orbit-averaged spectrum of the first generation of gamma rays in LS~5039 with an inclination $i=40\degr$, using the semi-analytical approach for various magnetic field intensity. The comparison of the expected flux in the 2-10~keV band with the recent {\em Suzaku} observations \citep{2009ApJ...697..592T} constrains the (uniform) magnetic field strength below 10~G. This result is in agreement with the development of a three-dimensional cascade (see Sect.~2). The one-generation approximation for the cascade is good in this case since for high magnetic field ($B>5~$G), the contribution from extra-generations can be ignored (see Sect.~\ref{sect_mag}). Note that the synchrotron peak energy emitted by secondary pairs barely changes with increasing magnetic field ($\epsilon_1\approx 1~$MeV, see Fig.~\ref{mag_orb}). This is due to the effect of synchrotron losses on the cooled energy distribution of the radiating pairs in the cascade. Synchrotron cooling dominates over Compton cooling ($t_{syn}<t_{ic}$) at high energies and depletes the most energetic pairs in the steady-state distribution (see Eq.~\ref{cool}). In consequence, the mean energy of cooled pairs in the cascade diminishes with increasing magnetic field (for a fixed stellar radiation field). The non-trivial combination of both effects results in a (almost) constant synchrotron peak (the critical energy in synchrotron radiation is proportional to $\gamma_e^2 B$).

\section{Conclusion}

Three-dimensional pair cascade can be initiated in gamma-ray binaries provided that pairs are confined and isotropized by the ambient magnetic field in the system. In LS~5039, a three-dimensional pair cascade contributes significantly in the formation of the VHE radiation at every orbital phase. In particular, the cascade radiation prevails over the primary source of gamma rays close to superior conjunction ({\em i.e.} where the $\gamma\gamma$-opacity is high) and gives a lower flux than the 1D cascade at this phase. The 3D cascade radiation is modulated differently compared with the 1D cascade and preserves the modulation of the primary absorbed flux because the pairs stay localized. In addition, the 3D cascade radiation decreases the amplitude of the observed TeV modulation. The amplitude of the HESS light curve is correctly reproduced for an inclination of $i\approx 40\degr$.

The ambient magnetic field in LS~5039 cannot exceed 10~G (if uniform) or synchrotron radiation from pairs in the cascade would overestimate X-ray observations. This is a reasonable constraint as most of massive stars are probably non-magnetic, even though strong magnetic fields ($>100~$G) have been measured for a few O stars at their surface (see \citealt{2009ARA&A..47..333D} for a recent review and references therein). The VHE emitter should also remain very close to the compact object location, possibly at the collision site between both star winds, otherwise the TeV light curve shape is not reproduced although this does not rule out complex combinations.

The model described in this paper is not fully satisfying. The spectral shape of VHE gamma rays is still not reproduced close to superior conjunction. In addition, the light curve amplitude tends to be overestimated except for low inclinations but then the shape is not perfect. It remains difficult to explain both the shape and the amplitude of the modulation in LS~5039. A possible solution would be to consider a more complex injection of fresh pairs along the orbit or additional effects such as adiabatic losses or advection. A Doppler-boosted emission in the primary source can also change the spectrum seen by the observer, especially around superior conjunction \citep{2010arXiv1004.0511D}. The primary source of gamma rays might be extended, VHE photons would come from {\em e.g.} the shock front between the pulsar wind and the stellar wind or along a relativistic jet. The development of an anisotropic 3D cascade is not excluded as well. Nevertheless, the calculations show that a three dimensional pair cascading provides a plausible framework to understand the TeV modulation in LS 5039.

\begin{acknowledgements}
This work was supported by the {\em European Community} via contract ERC-StG-200911.
\end{acknowledgements}

\bibliographystyle{aa}
\bibliography{cas3d}

\begin{thebibliography}{31}
\expandafter\ifx\csname natexlab\endcsname\relax\def\natexlab#1{#1}\fi

\bibitem[{{Abdo} {et~al.}(2009){Abdo}, {Ackermann}, {Ajello}, {Atwood},
  {Axelsson}, {Baldini}, {Ballet}, {Barbiellini}, {Bastieri}, {Baughman},
  {Bechtol}, {Bellazzini}, {Berenji}, {Blandford}, {Bloom}, {Bonamente},
  {Borgland}, {Bregeon}, {Brez}, {Brigida}, {Bruel}, {Burnett}, {Buson},
  {Caliandro}, {Cameron}, {Caraveo}, {Casandjian}, {Cavazzuti}, {Cecchi}, {{\c
  C}elik}, {Chaty}, {Chekhtman}, {Cheung}, {Chiang}, {Ciprini}, {Claus},
  {Cohen-Tanugi}, {Cominsky}, {Conrad}, {Corbel}, {Corbet}, {Cutini}, {Dermer},
  {de Angelis}, {de Palma}, {Digel}, {Silva}, {Drell}, {Dubois}, {Dubus},
  {Dumora}, {Farnier}, {Favuzzi}, {Fegan}, {Focke}, {Fortin}, {Frailis},
  {Fukazawa}, {Funk}, {Fusco}, {Gargano}, {Gasparrini}, {Gehrels}, {Germani},
  {Giebels}, {Giglietto}, {Giordano}, {Glanzman}, {Godfrey}, {Grenier},
  {Grondin}, {Grove}, {Guillemot}, {Guiriec}, {Hanabata}, {Harding},
  {Hayashida}, {Hays}, {Hill}, {Horan}, {Hughes}, {Jackson}, {J{\'o}hannesson},
  {Johnson}, {Johnson}, {Johnson}, {Kamae}, {Katagiri}, {Kataoka}, {Kawai},
  {Kerr}, {Kn{\"o}dlseder}, {Kocian}, {Kuehn}, {Kuss}, {Lande}, {Larsson},
  {Latronico}, {Lemoine-Goumard}, {Longo}, {Loparco}, {Lott}, {Lovellette},
  {Lubrano}, {Madejski}, {Makeev}, {Marelli}, {Mazziotta}, {Mc Enery},
  {Meurer}, {Michelson}, {Mitthumsiri}, {Mizuno}, {Moiseev}, {Monte},
  {Monzani}, {Morselli}, {Moskalenko}, {Murgia}, {Nolan}, {Norris}, {Nuss},
  {Ohsugi}, {Omodei}, {Orlando}, {Ormes}, {Ozaki}, {Paneque}, {Panetta},
  {Parent}, {Pelassa}, {Pepe}, {Pesce-Rollins}, {Piron}, {Porter}, {Rain{\`o}},
  {Rando}, {Ray}, {Razzano}, {Rea}, {Reimer}, {Reimer}, {Reposeur}, {Ritz},
  {Rochester}, {Rodriguez}, {Romani}, {Roth}, {Ryde}, {Sadrozinski}, {Sanchez},
  {Sander}, {Saz Parkinson}, {Scargle}, {Sgr{\`o}}, {Sierpowska-Bartosik},
  {Siskind}, {Smith}, {Smith}, {Spandre}, {Spinelli}, {Strickman}, {Suson},
  {Tajima}, {Takahashi}, {Takahashi}, {Tanaka}, {Tanaka}, {Thayer}, {Thompson},
  {Tibaldo}, {Torres}, {Tosti}, {Tramacere}, {Uchiyama}, {Usher}, {Vasileiou},
  {Venter}, {Vilchez}, {Vitale}, {Waite}, {Wallace}, {Wang}, {Winer}, {Wood},
  {Ylinen}, \& {Ziegler}}]{2009ApJ...706L..56A}
{Abdo}, A.~A., {Ackermann}, M., {Ajello}, M., {et~al.} 2009, \apjl, 706, L56

\bibitem[{{Aharonian} {et~al.}(2005){Aharonian}, {Akhperjanian}, {Aye},
  {Bazer-Bachi}, {Beilicke}, {Benbow}, {Berge}, {Berghaus}, {Bernl{\"o}hr},
  {Boisson}, {Bolz}, {Borrel}, {Braun}, {Breitling}, {Brown}, {Gordo},
  {Chadwick}, {Chounet}, {Cornils}, {Costamante}, {Degrange}, {Dickinson},
  {Djannati-Ata{\"i}}, {Drury}, {Dubus}, {Emmanoulopoulos}, {Espigat},
  {Feinstein}, {Fleury}, {Fontaine}, {Fuchs}, {Funk}, {Gallant}, {Giebels},
  {Gillessen}, {Glicenstein}, {Goret}, {Hadjichristidis}, {Hauser},
  {Heinzelmann}, {Henri}, {Hermann}, {Hinton}, {Hofmann}, {Holleran}, {Horns},
  {Jacholkowska}, {de Jager}, {Kh{\'e}lifi}, {Komin}, {Konopelko}, {Latham},
  {Le Gallou}, {Lemi{\`e}re}, {Lemoine-Goumard}, {Leroy}, {Lohse}, {Marcowith},
  {Martin}, {Martineau-Huynh}, {Masterson}, {McComb}, {de Naurois}, {Nolan},
  {Noutsos}, {Orford}, {Osborne}, {Ouchrif}, {Panter}, {Pelletier}, {Pita},
  {P{\"u}hlhofer}, {Punch}, {Raubenheimer}, {Raue}, {Raux}, {Rayner}, {Reimer},
  {Reimer}, {Ripken}, {Rob}, {Rolland}, {Rowell}, {Sahakian}, {Saug{\'e}},
  {Schlenker}, {Schlickeiser}, {Schuster}, {Schwanke}, {Siewert}, {Sol},
  {Spangler}, {Steenkamp}, {Stegmann}, {Tavernet}, {Terrier}, {Th{\'e}oret},
  {Tluczykont}, {Vasileiadis}, {Venter}, {Vincent}, {V{\"o}lk}, \&
  {Wagner}}]{2005Sci...309..746A}
{Aharonian}, F., {Akhperjanian}, A.~G., {Aye}, K.-M., {et~al.} 2005, Science,
  309, 746

\bibitem[{{Aharonian} {et~al.}(2006){Aharonian}, {Akhperjanian}, {Bazer-Bachi},
  {Beilicke}, {Benbow}, {Berge}, {Bernl{\"o}hr}, {Boisson}, {Bolz}, {Borrel},
  {Braun}, {Brown}, {B{\"u}hler}, {B{\"u}sching}, {Carrigan}, {Chadwick},
  {Chounet}, {Cornils}, {Costamante}, {Degrange}, {Dickinson},
  {Djannati-Ata{\"i}}, {O'C.~Drury}, {Dubus}, {Egberts}, {Emmanoulopoulos},
  {Espigat}, {Feinstein}, {Ferrero}, {Fiasson}, {Fontaine}, {Funk}, {Funk},
  {F{\"u}{\ss}ling}, {Gallant}, {Giebels}, {Glicenstein}, {Goret},
  {Hadjichristidis}, {Hauser}, {Hauser}, {Heinzelmann}, {Henri}, {Hermann},
  {Hinton}, {Hoffmann}, {Hofmann}, {Holleran}, {Horns}, {Jacholkowska}, {de
  Jager}, {Kendziorra}, {Kh{\'e}lifi}, {Komin}, {Konopelko}, {Kosack},
  {Latham}, {Le Gallou}, {Lemi{\`e}re}, {Lemoine-Goumard}, {Lohse}, {Martin},
  {Martineau-Huynh}, {Marcowith}, {Masterson}, {Maurin}, {McComb}, {Moulin},
  {de Naurois}, {Nedbal}, {Nolan}, {Noutsos}, {Orford}, {Osborne}, {Ouchrif},
  {Panter}, {Pelletier}, {Pita}, {P{\"u}hlhofer}, {Punch}, {Raubenheimer},
  {Raue}, {Rayner}, {Reimer}, {Reimer}, {Ripken}, {Rob}, {Rolland}, {Rowell},
  {Sahakian}, {Santangelo}, {Saug{\'e}}, {Schlenker}, {Schlickeiser},
  {Schr{\"o}der}, {Schwanke}, {Schwarzburg}, {Shalchi}, {Sol}, {Spangler},
  {Spanier}, {Steenkamp}, {Stegmann}, {Superina}, {Tavernet}, {Terrier},
  {Tluczykont}, {van Eldik}, {Vasileiadis}, {Venter}, {Vincent}, {V{\"o}lk},
  {Wagner}, \& {Ward}}]{2006A&A...460..743A}
{Aharonian}, F., {Akhperjanian}, A.~G., {Bazer-Bachi}, A.~R., {et~al.} 2006,
  \aap, 460, 743

\bibitem[{{Aragona} {et~al.}(2009){Aragona}, {McSwain}, {Grundstrom}, {Marsh},
  {Roettenbacher}, {Hessler}, {Boyajian}, \& {Ray}}]{2009ApJ...698..514A}
{Aragona}, C., {McSwain}, M.~V., {Grundstrom}, E.~D., {et~al.} 2009, \apj, 698,
  514

\bibitem[{{Bednarek}(1997)}]{1997A&A...322..523B}
{Bednarek}, W. 1997, \aap, 322, 523

\bibitem[{{Bednarek}(2006)}]{2006MNRAS.368..579B}
{Bednarek}, W. 2006, \mnras, 368, 579

\bibitem[{{Bednarek}(2007)}]{2007A&A...464..259B}
{Bednarek}, W. 2007, \aap, 464, 259

\bibitem[{{Blumenthal} \& {Gould}(1970)}]{1970RvMP...42..237B}
{Blumenthal}, G.~R. \& {Gould}, R.~J. 1970, Reviews of Modern Physics, 42, 237

\bibitem[{{Bonometto} \& {Rees}(1971)}]{1971MNRAS.152...21B}
{Bonometto}, S. \& {Rees}, M.~J. 1971, \mnras, 152, 21

\bibitem[{{Bosch-Ramon} {et~al.}(2008{\natexlab{a}}){Bosch-Ramon},
  {Khangulyan}, \& {Aharonian}}]{2008A&A...482..397B}
{Bosch-Ramon}, V., {Khangulyan}, D., \& {Aharonian}, F.~A. 2008{\natexlab{a}},
  \aap, 482, 397

\bibitem[{{Bosch-Ramon} {et~al.}(2008{\natexlab{b}}){Bosch-Ramon},
  {Khangulyan}, \& {Aharonian}}]{2008A&A...489L..21B}
{Bosch-Ramon}, V., {Khangulyan}, D., \& {Aharonian}, F.~A. 2008{\natexlab{b}},
  \aap, 489, L21

\bibitem[{{B{\"o}ttcher} \& {Dermer}(2005)}]{2005ApJ...634L..81B}
{B{\"o}ttcher}, M. \& {Dermer}, C.~D. 2005, \apjl, 634, L81

\bibitem[{{B\"{o}ttcher} \& {Schlickeiser}(1997)}]{1997A&A...325..866B}
{B\"{o}ttcher}, M. \& {Schlickeiser}, R. 1997, \aap, 325, 866

\bibitem[{{Casares} {et~al.}(2005){Casares}, {Rib{\'o}}, {Ribas}, {Paredes},
  {Mart{\'{\i}}}, \& {Herrero}}]{2005MNRAS.364..899C}
{Casares}, J., {Rib{\'o}}, M., {Ribas}, I., {et~al.} 2005, \mnras, 364, 899

\bibitem[{{Cerutti} {et~al.}(2009{\natexlab{a}}){Cerutti}, {Dubus}, \&
  {Henri}}]{2009arXiv0912.3722C}
{Cerutti}, B., {Dubus}, G., \& {Henri}, G. 2009{\natexlab{a}}, ArXiv e-prints

\bibitem[{{Cerutti} {et~al.}(2009{\natexlab{b}}){Cerutti}, {Dubus}, \&
  {Henri}}]{2009A&A...507.1217C}
{Cerutti}, B., {Dubus}, G., \& {Henri}, G. 2009{\natexlab{b}}, \aap, 507, 1217

\bibitem[{{Donati} \& {Landstreet}(2009)}]{2009ARA&A..47..333D}
{Donati}, J. \& {Landstreet}, J.~D. 2009, \araa, 47, 333

\bibitem[{{Dubus}(2006{\natexlab{a}})}]{2006A&A...451....9D}
{Dubus}, G. 2006{\natexlab{a}}, \aap, 451, 9

\bibitem[{{Dubus}(2006{\natexlab{b}})}]{2006A&A...456..801D}
{Dubus}, G. 2006{\natexlab{b}}, \aap, 456, 801

\bibitem[{{Dubus} {et~al.}(2008){Dubus}, {Cerutti}, \&
  {Henri}}]{2008A&A...477..691D}
{Dubus}, G., {Cerutti}, B., \& {Henri}, G. 2008, \aap, 477, 691

\bibitem[{{Dubus} {et~al.}(2010){Dubus}, {Cerutti}, \&
  {Henri}}]{2010arXiv1004.0511D}
{Dubus}, G., {Cerutti}, B., \& {Henri}, G. 2010, ArXiv e-prints

\bibitem[{{Ginzburg} \& {Syrovatskii}(1964)}]{1964ocr..book.....G}
{Ginzburg}, V.~L. \& {Syrovatskii}, S.~I. 1964, {The Origin of Cosmic Rays},
  ed. V.~L. {Ginzburg} \& S.~I. {Syrovatskii}

\bibitem[{{Maraschi} \& {Treves}(1981)}]{1981MNRAS.194P...1M}
{Maraschi}, L. \& {Treves}, A. 1981, \mnras, 194, 1P

\bibitem[{{Martocchia} {et~al.}(2005){Martocchia}, {Motch}, \&
  {Negueruela}}]{2005A&A...430..245M}
{Martocchia}, A., {Motch}, C., \& {Negueruela}, I. 2005, \aap, 430, 245

\bibitem[{{Mastichiadis}(1991)}]{1991MNRAS.253..235M}
{Mastichiadis}, A. 1991, \mnras, 253, 235

\bibitem[{{McSwain} {et~al.}(2004){McSwain}, {Gies}, {Huang}, {Wiita},
  {Wingert}, \& {Kaper}}]{2004ApJ...600..927M}
{McSwain}, M.~V., {Gies}, D.~R., {Huang}, W., {et~al.} 2004, \apj, 600, 927

\bibitem[{{Motch} {et~al.}(1997){Motch}, {Haberl}, {Dennerl}, {Pakull}, \&
  {Janot-Pacheco}}]{1997A&A...323..853M}
{Motch}, C., {Haberl}, F., {Dennerl}, K., {Pakull}, M., \& {Janot-Pacheco}, E.
  1997, \aap, 323, 853

\bibitem[{{Sierpowska} \& {Bednarek}(2005)}]{2005MNRAS.356..711S}
{Sierpowska}, A. \& {Bednarek}, W. 2005, \mnras, 356, 711

\bibitem[{{Sierpowska-Bartosik} \& {Torres}(2008)}]{2008APh....30..239S}
{Sierpowska-Bartosik}, A. \& {Torres}, D.~F. 2008, Astroparticle Physics, 30,
  239

\bibitem[{{Takahashi} {et~al.}(2009){Takahashi}, {Kishishita}, {Uchiyama},
  {Tanaka}, {Yamaoka}, {Khangulyan}, {Aharonian}, {Bosch-Ramon}, \&
  {Hinton}}]{2009ApJ...697..592T}
{Takahashi}, T., {Kishishita}, T., {Uchiyama}, Y., {et~al.} 2009, \apj, 697,
  592

\bibitem[{{Zdziarski} {et~al.}(2009){Zdziarski}, {Malzac}, \&
  {Bednarek}}]{2009MNRAS.tmpL.175Z}
{Zdziarski}, A.~A., {Malzac}, J., \& {Bednarek}, W. 2009, \mnras, L175+

\end{thebibliography}

\end{document}